\documentclass[10pt,aip,jap,showpacs,amsfonts,amsmath,amssymb,twocolumn,floatfix,superscriptaddress]{revtex4-1}
\usepackage[final]{graphicx}
\usepackage{epstopdf}
\usepackage{color}

\newcommand{\bs}{\boldsymbol}
\newcommand{\Gr}{{\stackrel{\leftrightarrow}{G}}}
\newcommand{\im}{\operatorname{Im}}
\newcommand{\re}{\operatorname{Re}}

\begin{document}

\title{Fluctuational electrodynamics of hyperbolic metamaterials}


\author{Yu Guo}
\affiliation{Department of Electrical and Computer Engineering,
University of Alberta, Edmonton, AB, T6G 2V4, Canada}

\author{Zubin Jacob}
\affiliation{Department of Electrical and Computer Engineering,
University of Alberta, Edmonton, AB, T6G 2V4, Canada}

\begin{abstract} 
We give a detailed account of equilibrium and non-equilibrium fluctuational electrodynamics of hyperbolic metamaterials. We show the unifying aspects of two different approaches; one utilizes the second kind of fluctuation dissipation theorem and the other makes use of the scattering method. We analyze the near-field of hyperbolic media at finite temperatures and show that the lack of spatial coherence can be attributed to the multi-modal nature of super-Planckian thermal emission.  We also adopt the analysis to phonon-polaritonic super-lattice metamaterials and describe the regimes suitable for experimental verification of our predicted effects. The results reveal that far-field thermal emission spectra are dominated by epsilon-near-zero and epsilon-near-pole responses as expected from Kirchoff's laws. Our work should aid both theorists and experimentalists to study complex media and engineer equilibrium and non-equilibrium fluctuations for applications in thermal photonics.
\end{abstract}

\maketitle

\section{Introduction}
The foundations of analyzing thermal and vacuum fluctuations of the electromagnetic field inside matter were laid in the seminal work of S.M.Rytov\cite{rytov_principles_1989}. This later gave rise to a unified approach of understanding fluctuational forces\cite{lifshitz_theory_1956} (Lifshitz theory of Casimir forces), near field thermal emission and radiative heat transfer\cite{polder_theory_1971,zhang_nano/microscale_2007,chen_nanoscale_2005,biehs_nanoscale_2011,shchegrov_near-field_2000,guo_broadband_2012,joulain_surface_2005,laroche_near-field_2006,lussange_radiative_2012,nefedov_giant_2011,mulet_enhanced_2002,narayanaswamy_thermal_2004,rousseau_radiative_2009,volokitin_radiative_2001,volokitin_near-field_2007,pendry_radiative_1999,joulain_noncontact_2010,pendry_radiative_1999,vinogradov_thermally_2009,fu_nanoscale_2006,francoeur_near-field_2008,otey_thermal_2010,francoeur_near-field_2008,hu_near-field_2008,shen_surface_2009,liu_taming_2011}. (Polder-Van-Hove theory\cite{polder_theory_1971}). Recent developments in nanoengineering and detection have led to experimental regimes\cite{guha_near-field_2012,de_wilde_thermal_2006,hu_near-field_2008,narayanaswamy_surface_2003,shen_surface_2009,jones_thermal_2012,liu_taming_2011} where these effects can play a dominant role. Simultaneously, theoretical work has shed light on the fact that the classical scattering matrix along with the temperatures of objects of various geometries can completely characterize these fluctuations in both equilibrium and non-equilibrium situations\cite{bimonte_scattering_2009,kruger_nonequilibrium_2011,kruger_trace_2012,maghrebi_scattering_2013,messina_casimir-lifshitz_2011,messina_scattering-matrix_2011,rahi_scattering_2009,antezza_casimir-lifshitz_2006,bimonte_general_2007,bimonte_theory_2006,reid_fluctuation-induced_2013,rodriguez_fluctuating-surface-current_2012,rodriguez_fluctuating-surface-current_2013}.

Metamaterials are artificial media designed to achieve exotic electromagnetic responses that are beyond those available in conventional materials\cite{engheta_metamaterials:_2006,shalaev_optical_2007,smith_metamaterials_2004}. A large body of work has emerged in the last decade which in principle engineers the classical scattering matrix to achieve effects such as negative refraction\cite{pendry_negative_2000,veselago_electrodynamics_1968}, enhanced chirality\cite{plum_metamaterial_2009,wang_chiral_2009,zhang_negative_2009,pendry_negative_2000,veselago_electrodynamics_1968}, invisibility\cite{cai_optical_2007,pendry_controlling_2006,schurig_metamaterial_2006} and subwavelength imaging\cite{fang_subdiffraction-limited_2005,cai_optical_2007,pendry_controlling_2006,schurig_metamaterial_2006,jacob_optical_2006}. Recently, it was shown that a specific class of metamaterials, known as hyperbolic media\cite{cortes_quantum_2012,guo_applications_2012,jacob_optical_2006,krishnamoorthy_topological_2012,smith_electromagnetic_2003,smith_negative_2004,podolskiy_strongly_2005}(indefinite media) has the potential for thermal engineering. Such media support unique modes which can be thermally excited and detected in the near-field due to the super-Planckian nature of their thermal emission spectrum\cite{biehs_hyperbolic_2012,guo_broadband_2012,guo_thermal_2013,simovski_optimization_2013,biehs_super-planckian_2013}. 

In this paper, we adopt the techniques of fluctuational electrodynamics to provide a first-principle account of the thermal emission characteristics of hyperbolic media. We show that the conventional approach of utilizing the second kind of fluctuation dissipation theorem\cite{rytov_principles_1989,eckhardt_first_1982,eckhardt_macroscopic_1984} is equivalent to the scattering matrix method\cite{bimonte_general_2007,bimonte_scattering_2009,eckhardt_first_1982,eckhardt_macroscopic_1984} for calculating the metamaterial energy density. We specifically provide the derivations of the fluctuational effects in both effective medium theory and practical thin film multilayer metamaterial designs\cite{kidwai_effective-medium_2012,cortes_quantum_2012}. While the characteristics can in principle be obtained from formulas related to the reflection coefficients, it does not shed light on various aspects of equilibrium or non-equilibrium fluctuations in the context of metamaterials. Our aim is to provide an insightful look at prevailing approaches adopted to the case of hyperbolic media. 

We also consider the case of a practical phonon-polaritonic metamaterial\cite{guo_broadband_2012,korobkin_measurements_2010} and show the stark contrast in the far-field and near-field thermal emission characteristics\cite{shchegrov_near-field_2000}. This should help experimentalists design experiments starting from analyzing the far-field characteristics, retrieving effective medium characteristics and then look for our predicted near-field effects. We show that the far-field characteristics are dominated by the epsilon-near-zero and epsilon-near-pole responses as expected from Kirchoff's laws\cite{molesky_high_2013}. This is true independent of material choice and can occur for both nanowire and multilayer hyperbolic media\cite{molesky_high_2013}. We comment here that for practical applications high temperature plasmonics and metamaterials would be needed\cite{molesky_high_2013}. 

We also study the limitations of effective medium theory (EMT) but focus on cases where there is good agreement between practical structures and EMT\cite{tschikin_limits_2013,cortes_quantum_2012,kidwai_effective-medium_2012}. We emphasize that it is known in the metamaterials community that the unit cell of a metamaterial can show characteristics similar to the bulk medium\cite{cortes_quantum_2012}. In the context of thin film hyperbolic media, this was experimentally elucidated in Ref.~\onlinecite{kim_improving_2012} and theoretically explained in detail in Ref.~\onlinecite{cortes_quantum_2012}.

In this paper we also describe another effect connected to hyperbolic super-Planckian thermal emission\cite{guo_broadband_2012}. We analyze the spatial coherence\cite{carminati_near-field_1999,henkel_spatial_2000,joulain_near_2008,joulain_surface_2005,lau_spatial_2007} of the near-field thermal emission and relate it to the metamaterial modes. We show that there is a subtle interplay in near-field spatial coherence due to competition between surface waves and hyperbolic modes. We expect our work to aid experimentalists in isolating thermal effects related to metamaterials and also form the theoretical foundation for developing the macroscopic quantum electrodynamics\cite{scheel_macroscopic_2008} of hyperbolic media.

\section{Fluctuation dissipation theorem}
In global thermal equilibrium, the first kind of fluctuation dissipation theorem\cite{rytov_principles_1989,eckhardt_macroscopic_1984}(FDT) directly specifies the correlation function of electric fields. It is expressed by

\begin{align}
\label{first kind}
\left\langle{\vec E(\bs r_1,\omega)\otimes \vec E^*(\bs r_2,\omega')}\right\rangle
&= \nonumber \\
\frac{\mu_0\omega}{\pi}\Theta(\omega,T)&\im\Gr(\bs r_1,\bs r_2,\omega)\delta(\omega-\omega').
\end{align}
Here  $\Gr$ is the dyadic Green's function\cite{tai_dyadic_1994,kong_electromagnetic_1990}(DGF), $\Theta(\omega,T)=\hbar\omega/(e^{\hbar\omega/{k_BT}}-1)$ is the mean energy of a thermal oscillator. 

Eq. (\ref{first kind}) has two main applications. Firstly, it can be used to derive the electromagnetic stress tensor at a certain point. Secondly, it directly gives the cross-spectral density tensor\cite{carminati_near-field_1999,lau_spatial_2007} which characterizes the spatial coherence of a thermal radiative source. The second kind of FDT\cite{rytov_principles_1989,eckhardt_macroscopic_1984} that specifies the correlation function of thermally generated random currents is 
\begin{align}
&\left\langle{\vec{j}(\bs r_1,\omega)\otimes\vec{j}^*(\bs r_2,\omega')}\right\rangle=\nonumber \\
&\frac{\omega \epsilon_0}{\pi} \bs \epsilon''(\omega)\Theta(\omega,T)\delta(\bs r_1-\bs r_2)\delta(\omega-\omega'). 
\end{align}
We assume the permittivity $\bs \epsilon$ is a diagonal matrix;  $\bs \epsilon''$ denotes the imaginary part. 

The first kind of FDT can only be used in global thermal equilibrium. In non-equilibrium situation, we should first employ Maxwell equations to obtain the electromagnetic fields generated by random currents through the DGF, 
\begin{align}
&\vec E(\bs r) = i\omega {\mu _0}\iiint {\Gr (\bs r,\bs r')\vec j(\bs r')}d\bs r', \\
&\vec H(\bs r)=\iiint  \nabla \times \Gr (\bs r, \bs r')\vec j(\bs r')d \bs r',
\end{align}
and then calculate the electromagnetic stress tensor or the cross-spectral density tensor. 

The dyadic Green's function (DGF) satisfies an important identity\cite{eckhardt_macroscopic_1984,novotny_principles_2006}, 
\begin{align}
&\im \Gr (\bs r_1, \bs r_2,\omega)=\nonumber \\
&\frac{\omega^2}{c^2}\int_V \Gr (\bs r_1,\bs r',\omega)\bs \epsilon''(\bs r',\omega)\Gr^\dag(\bs r_2,\bs r',\omega)d^3 \bs r'.
\end{align}
This identity ensures that at global thermal equilibrium the first kind and the second kind of FDT lead to identical results. 

\section{Thermal emission from half space uniaxial media}
In this section, we consider an uniaxial medium located in the lower space $(z<0)$ at temperature T while the upper space vacuum part is at zero temperature. The relative permittivity of the uniaxial medium is a diagonal matrix, $\bs{\epsilon}=diag[\epsilon_{\parallel};\epsilon_{\parallel};\epsilon_{\perp}]$. Note that hyperbolic metamaterials are a special kind of uniaxial medium satisfying $\epsilon_{\parallel}\epsilon_{\perp}<0$. As mentioned before, we should employ the second kind of FDT because this is a non-equilibrium problem. 

To solve DGF in planar structures, it is convenient to work in the wavevector space. DGF in vacuum\cite{kong_electromagnetic_1990} is ($z>z'$)
\begin{align}
\Gr(\bs r,\bs r',\omega) =& \frac{i}{{8{\pi ^2}}}\iint \frac{{d{k_x}d{k_y}}}{k_{z0}}{e^{i{\bs k_\perp}\cdot(\bs r_\perp^{} -\bs r_\perp^\prime)}} \nonumber \\
&\{{{\hat s}_ + ^0 }{{\hat s}_ +^0 }{e^{i{k_{z0}}(z - z')}} + {{\hat p}_ +^0 }{{\hat p}_ + ^0}{e^{i{k_{z0}}(z - z')}}\}
\end{align}
Here we define ${\hat k_ + } = ({k_x},{k_y},{k_{z0}})/k_0$ is the normalized wave-vector of upward waves ($z > z'$) in free space, 
 $\bs k_\perp=(k_x,k_y)$, ${k_\rho} = \sqrt {k_x^2 + k_y^2}$, $k_{z0} = \sqrt {k_0^2 - k_\rho ^2}$, and $\bs r_\perp=(x,y)$.  ${\hat s_+^0 } = {{{\hat k}_ + } \times \hat z} = {({k_y}, - {k_x},0)} / {k_\rho}$ is the unit direction vector of s-polarized waves, ${\hat p_ +^0} = {\hat s_ + ^0} \times {\hat k_ + } = {{( - {k_x}{k_{z0}}, - {k_y}{k_{z0}},k_\rho ^2)}/{{k_0}{k_\rho }}}$ is the unit direction vector of p-polarized waves. Correspondingly and for later use, ${\hat k_ - } = ({k_x},{k_y},-{k_{z0}})/{k_0}$ is the normalized wave-vector of downward waves (when $z < z'$), ${\hat s_ - ^0 } = {{{\hat k}_ - } \times \hat z} = {({k_y}, - {k_x},0)} / {k_\rho}$ same with ${\hat s_ + ^0}$ , and ${\hat p_ -^0 } = {\hat s_ -^0 } \times {\hat k_ -} = {{({k_x}{k_{z0}},{k_y}{k_{z0}},k_\rho ^2)} / {{k_0}{k_\rho }}}$. 
 
The DGF relating thermally generated random currents inside the medium in the lower space to the fields in upper space vacuum is
\begin{align}
&\Gr_{01}(\bs r,\bs r') = \frac{i}{{8{\pi ^2}}}\iint \frac{{d{k_x}d{k_y}}}{{{k_{z0}}}}{e^{i{\bs k_\perp}\cdot(\bs r_\perp^{} -\bs r_\perp^\prime)}} \nonumber \\
& \{ {t^s}{{\hat s}_ + ^0 }{{\hat s}_+^1 }{e^{i{k_{z0}}z-i{k_{zs}}z'}} + {t^p}{{\hat p}_ +^0 }{{\hat p}_ +^1 }{e^{i{k_{z0}}z-i{k_{zp}}z'}} \}. 
\end{align}
Here, ${k_{zs}} = \sqrt {\epsilon_\parallel k_0^2 - k_\rho ^2}$, ${k_{zp}} = \sqrt {\epsilon_\parallel k_0^2 - \frac{\epsilon_\parallel}{\epsilon_\perp} k_\rho ^2}$. $\hat s_+^1=\hat s_+^0$, and ${\hat p_ + ^1 } =  {{( -{k_x}{k_{zp}}, -{k_y}{k_{zp}},k_\rho ^2{\epsilon_\parallel}/{\epsilon_\perp})}/{{k_0}{k_\rho }}\sqrt{\epsilon_\parallel}}$  which are the unit direction vectors of s- and p-polarized waves inside the unaxial medium, respectively. 
Note the transmission coefficients incident from the vacuum side should be in terms of the electric fields, 
\begin{equation}
t^s=\frac{2k_{z0}}{k_{z0}+k_{zs}},\quad t^p=\frac{2k_{z0}\sqrt{\epsilon_{\parallel}}}{\epsilon_{\parallel}k_{z0}+k_{zp}}.
\end{equation}

To calculate the magnetic fields, we should evaluate $\nabla\times\Gr_{01}$, which can be easily done in the wavevector space. The curl operator will work on the first vector of $\Gr_{01}$, 
\begin{align}
&\nabla\times\Gr_{01}(\bs r,\bs r')=\frac{k_0}{{8{\pi ^2}}}\iint \frac{{d{k_x}d{k_y}}}{{{k_{z0}}}}{e^{i{\bs k_\perp}\cdot(\bs r_\perp^{} -\bs r_\perp^\prime)}}  \nonumber \\
&\{{t^s}{{\hat p}_ + ^0 }{{\hat s}_+^1 }{e^{i{k_{z0}}z-i{k_{zs}}z'}} - {t^p}{{\hat s}_ +^0 }{{\hat p}_ +^1 }{e^{i{k_{z0}}z-i{k_{zp}}z'}}\}.
\end{align}

The free space energy density is defined by 
\begin{align}
u(\omega,\bs r)=2\left(\frac{1}{2}\epsilon_0\operatorname {Tr}\left\langle{\vec E(\omega,\bs r)\otimes \vec E^*(\omega,\bs r)}\right\rangle \right. \nonumber \\
\left. +\frac{1}{2}\mu_0\operatorname {Tr}\left\langle{\vec H(\omega,\bs r)\otimes \vec H^*(\omega,\bs r)}\right\rangle \right), 
\end{align}
where the prefactor 2 accounts for the negative frequency counterpart. 
Following the formalism in Ref.~\onlinecite{lau_spatial_2007}, we define 
\begin{align}
&g_{e}(\bs k_{\perp},z,z',\omega)=-\frac{1}{2k_{z0}} \nonumber \\
&\left\{ {{t^s}{{\hat s}_ + ^0 }{{\hat s}_ +^1 }{e^{ik_{z0}z-i{k_{zs}}z'}} + {t^p}{{\hat p}_ +^0 }{{\hat p}_ +^1 }{e^{ik_{z0}z-i{k_{zp}}z'}}} \right\},\\
&g_{h}(\bs k_{\perp},z,z',\omega)=\frac{1}{2k_{z0}} \nonumber \\
& \left\{{{t^s}{{\hat p}_ + ^0 }{{\hat s}_+^1 }{e^{ik_{z0}z-i{k_{zs}}z'}} - {t^p}{{\hat s}_ +^0 }{{\hat p}_ +^1 }{e^{ik_{z0}z-i{k_{zp}}z'}}}\right\}.
\end{align}
\begin{widetext}
One can then find
\begin{equation}
u(\omega,z)=\frac{{\omega}^3}{\pi c^4}\Theta(\omega,T)\int _{-\infty}^0 dz' \int  _{-\infty}^{+\infty} \frac{d^2 \bs k_{\perp}}{4{\pi}^2}
\left(  \operatorname {Tr}\left( g_{e}^{}\bs \epsilon'' g_{e}^\dagger \right) + \operatorname {Tr}\left( g_{h}^{}\bs \epsilon'' g_{h}^\dagger \right) \right)
\end{equation}
Inserting the expressions of $g_e$ and $g_h$, we have
\begin{align}
u(\omega,z)=\frac{{\omega}^3}{8\pi^2 c^4}\Theta(\omega,T)e^{-2\im(k_{z0})z}\int _{-\infty}^0 dz' \int  _{0}^{+\infty}k_{\rho}{dk_{\rho}}\frac{1}{|k_{z0}|^2}\left(1+\frac{k_{\rho}^2+|{k_{z0}^2|}}{k_0^2}\right) \nonumber \\
\left(\epsilon_{\parallel}''|t^s|^2e^{2\operatorname {Im}(k_{zs})z'}+\left(\dfrac{\epsilon_{\perp}''|\epsilon_{\parallel}/\epsilon_{\perp}|^2k_{\rho}^2+\epsilon_{\parallel}''|{k_{zp}^{2}|}}{k_0^2|\epsilon_{\parallel}^2|}\right)|t^p|^2e^{2\operatorname {Im}(k_{zp})z'}\right). 
\end{align}
\end{widetext}
The integration on $z'$ can be easily done. Further by taking the imaginary part of the dispersion relation 
\begin{equation}
\frac{k_\rho^2}{\epsilon_\parallel}+\frac{k_{zs}^2}{\epsilon_\parallel}=\frac{\omega^2}{c^2},\quad \frac{k_\rho^2}{\epsilon_\perp}+\frac{k_{zp}^2}{\epsilon_\parallel}=\frac{\omega^2}{c^2}
\end{equation}
for s- and p-polarized waves, this result can be simplified as
\begin{align}
\label{general expression}
&u(\omega,z) = \frac{{U{}_{BB}(\omega ,T)}}{2} \nonumber \\
&\left\{ {\int_0^{{k_0}} {\frac{{{k_\rho }d{k_\rho }}}{{{k_0}\left| {{k_{z0}}} \right|}}} } \right.\frac{{(1 - {{\left| {{r^s}} \right|}^2}) +(1- {{\left| {{r^p}} \right|}^2})}}{2} \nonumber \\
& + \left. {\int_{{k_0}}^\infty  {\frac{{k_\rho ^3d{k_\rho }}}{{k_0^3\left| {{k_{z0}}} \right|}}{e^{ - 2\operatorname{Im} ({k_{z0}})z}}(\operatorname{Im} ({r^s}) + \operatorname{Im} ({r^p}))} } \right\}. 
\end{align} 
Here $U_{BB}=\frac{\omega^2}{\pi^2 c^3}\Theta(\omega,T)$ is the energy density of blackbody.$r^s$ and $r^p$ are the standard reflection coefficients given by
\begin{equation}
r^s=\frac{k_{z0}-k_{zs}}{k_{z0}+k_{zs}},\quad r^p=\frac{\epsilon_{\parallel}k_{z0}-k_{zp}}{\epsilon_{\parallel}k_{z0}+k_{zp}}.
\end{equation}
The propagating wave part $ 1-|r|^2 $ in Eq.~(\ref{general expression}) is the far field emissivity, equivalent to Kirchhoff's law. Correspondingly, the evanescent wave part can be interpreted as Kirchhoff's law in the near field and $ 2\im(r) $ is the near field emissivity\cite{pendry_radiative_1999,biehs_mesoscopic_2010,guo_thermal_2013,mulet_enhanced_2002}, which is widely used in heat transfer problems. $ 2\im(r) $ is also proportional to the near field local density of states (LDOS) proposed in Ref.~\onlinecite{pendry_radiative_1999} and is related to the tunneling and subsequent absorption of energy carried by evanescent waves. Recently extensive theoretical and experimental works have demonstrated the ability of HMMs to enhance the near field LDOS\cite{cortes_quantum_2012,jacob_engineering_2010,krishnamoorthy_topological_2012}. Thus we expect the use of HMMs in thermal and energy management. 

\subsection{Energy in matter and fields}
We can use the above definitions to compare the energy density in the near-field of the hyperbolic media to any other control sample. A pertinent question is about how much energy density is in matter degrees of freedom as opposed to the fields. This is difficult to answer inside the medium but can be done unambiguously in the near-field. 

In the high-k approximation, where the wavevector parallel to the interface $k_\rho$ is sufficiently large, the near-field energy density is governed by the tunneling parameter which we define as the imaginary part of the p-polarized reflection coefficient. Thus studying the behavior of this tunneling parameter sheds light on the near-field energy density. In the low loss limit, the reflection for p-polarized waves incident on an interface between vacuum and HMM can be expressed by\cite{guo_broadband_2012,miller_effectiveness_2013}
\begin{equation}
\label{high_k_HMM}
\im(r_p^{\text{HMM}})\approx \frac{2\sqrt{|\epsilon_{\parallel}\epsilon_{\perp}|}}{1+|\epsilon_{\parallel}\epsilon_{\perp}|}.
\end{equation}
While for an isotropic medium, the high-k approximation gives 
\begin{equation}
\label{high_k_iso}
\im(r_p^{\text{iso}})\approx \frac{2\epsilon''}{|1+\epsilon|^2}.
\end{equation}
The most striking difference between the above equations is that for a conventional isotropic medium the near-field energy density is completely dominated by the imaginary part of the dielectric constant. These fluctuations disappear in the low loss limit and can be attributed to matter degrees of freedom. This is because the imaginary part of the dielectric constant which governs field fluctuations also characterizes the irreversible conversion of electromagnetic energy into thermal energy of matter degrees of freedom. On the other hand, the hyperbolic medium shows near-field fluctuations arising from high-k modes completely indpendent of material losses  and the energy resides in the field. 

Let us analyze what would happen at mid-infrared frequencies where phonon polaritonic materials can give rise to this low loss high-k limit for hyperbolic media. We clearly see from Eq.~\ref{high_k_iso} that the near field emissivity would be very small when the frequency is away from the surface phonon polariton resonance (SPhPR) frequency where $\re(\epsilon)=-1$. However, for HMMs made of phonon polaritonic materials and dielectrics, the near field emissivity (Eq.~\ref{high_k_HMM}) can be comparably large in broad frequency region, though in this approximation its magnitude cannot exceed one. Note here we do not account for surface wave resonances which can change the picture considerably especially if one wants to optimize near-field heat transfer\cite{miller_effectiveness_2013}. Our aim is to focus on the bulk modes only. 

\section{Thermal emission from multilayered structures}
In this section we will consider multilayered structures. In the field of metamaterials, multilayered structures are widely used to achieve effective uniaxial media. The aim here is to go beyond effective medium theory and calculate the exact thermal emission from multilayered structures using the second kind of FDT. We assume that the medium in all layers is isotropic and non-magneto-optical for simplicity. To find DGFs relating the random currents in each layer to the vacuum region, we will follow the method in Ref.~\onlinecite{kong_electromagnetic_1990}. Firstly assuming the current source is in the vacuum region, we can calculate the fields induced by the source in all the layers by transfer matrix method which matches the boundary conditions at all the interfaces. Thus the DGFs with source in the vacuum region are ready to be employed. Next we use the reciprocal property of the DGF to achieve DGF when the sources are in the lower space. 

DGF in the vacuum region ($z<z'$) is 
\begin{align}
&\Gr_{00}(\bs r,\bs r') = \frac{i}{{8{\pi ^2}}}\iint {\frac{{d{k_x}d{k_y}}}{k_{z0}}}{e^{i{\bs k_\perp}\cdot(\bs r_\perp^{} -\bs r_\perp^\prime)}} \nonumber \\
&\Big\{\left({\hat s}_-^0 e^{-ik_{z0}z}+r^s {\hat s}_+^0 e^{ik_{z0}z}\right){\hat s}_-^0 e^{ik_{z0}z'} \nonumber \\ 
&+ ({\hat p}_-^0 e^{-ik_{z0}z}+r^p {\hat p}_+^0 e^{ik_{z0}z}){\hat p}_-^0 e^{ik_{z0}z'}\Big\}
\end{align}
DGF in the intermediate slabs are
\begin{align}
&\Gr_{l0}(\bs r,\bs r') = \frac{i}{{8{\pi ^2}}}\iint {\frac{{d{k_x}d{k_y}}}{k_{z0}}}{e^{i{\bs k_\perp}\cdot(\bs r_\perp^{} -\bs r_\perp^\prime)}} \nonumber \\
&\Big\{\left(B_l{\hat s}_-^l e^{-ik_{zl}z}+A_l {\hat s}_+^l e^{ik_{zl}z}\right){\hat s}_-^0 e^{ik_{z0}z'} \nonumber \\
&+ (D_l{\hat p}_-^l e^{-ik_{zl}z}+C_l {\hat p}_+^l e^{ik_{zl}z}){\hat p}_-^0 e^{ik_{z0}z'}\Big\} 
\end{align}

DGF in the last layer is 
\begin{align}
&\Gr_{(N+1)0}(\bs r,\bs r') = \frac{i}{{8{\pi ^2}}}\iint {\frac{{d{k_x}d{k_y}}}{k_{z0}}}{e^{i{\bs k_\perp}\cdot(\bs r_\perp^{} -\bs r_\perp^\prime)}} \nonumber\\
&\left\{t_s{\hat s}_-^t e^{-ik_{zt}z}{\hat s}_-^0 e^{ik_{z0}z'} \right. 
+ \left.t_p{\hat p}_-^t e^{-ik_{zt}z}{\hat p}_-^0 e^{ik_{z0}z'}\right\}
\end{align}
Note in the last layer we only have the downward waves, namely, the transmission. 

The boundary conditions give\cite{kong_electromagnetic_1990}
\begin{align}
& A_l e^{ik_{zl}z_l}+B_l e^{-ik_{zl}z_l} =\nonumber\\
& A_{l+1} e^{ik_{z(l+1)}z_l}+B_{l+1} e^{-ik_{z(l+1)}z_l} \\
& k_{zl}(A_l e^{ik_{zl}z_l}-B_l e^{-ik_{zl}z_l}) =\nonumber \\
& k_{z(l+1)}(A_{l+1} e^{ik_{z(l+1)}z_l}-B_{l+1} e^{-ik_{z(l+1)}z_l})
\end{align}
for s-polarized waves, and 
\begin{align}
&\sqrt{\epsilon_l}(C_l e^{ik_{zl}z_l}+D_l e^{-ik_{zl}z_l})=\nonumber\\
&\sqrt{\epsilon_{l+1}}(C_{l+1} e^{ik_{z({l+1})}z_l}+D_{l+1} e^{-ik_{z(l+1)}z_l}) \\
&\frac{k_{zl}}{\sqrt{\epsilon_l}}(C_l e^{ik_{zl}z_l}-D_l e^{-ik_{zl}z_l}) = \nonumber\\
 &\frac{k_{z({l+1})}}{\sqrt{\epsilon_{l+1}}}(C_{l+1} e^{ik_{z(l+1)}z_l}-D_{l+1} e^{-ik_{z(l+1)}z_l})
\end{align}
for p-polarized waves. 
Following the same steps as in the uniaxial case, the final expression is 
\begin{widetext}
\begin{align}
u(\omega,z)=\frac{{\omega}^3}{8 \pi^2 c^4}\Theta(\omega,T)e^{-2\im(k_{z0})z}\sum \limits _{l=1}^{N+1}\int _{z_l}^{z_{l-1}} dz' \int_{0}^{+\infty}k_{\rho}{dk_{\rho}}\frac{1}{|k_{z0}|^2}\left(1+\frac{k_{\rho}^2+|{k_{z0}^2|}}{k_0^2}\right)\epsilon_{l}'' \nonumber \\
\left(\left |A_le^{ik_{zl}z'}+B_le^{-ik_{zl}z'}\right |^2+\left |\frac{k_{zl}(C_le^{ik_{zl}z'}-D_le^{-ik_{zl}z'})}{k_0\sqrt{\epsilon_l}}\right |^2+\left |\frac{k_{\rho}(C_le^{ik_{zl}z'}+D_le^{-ik_{zl}z'})}{k_0\sqrt{\epsilon_l}}\right |^2 \right), 
\end{align}
\end{widetext}
where $N$ is the total number of layers in the structure. 

To simplify the above result, we first note that the integral 
\begin{align}
\label{last layer}
&\int _{z_l}^{z_{l-1}} dz' k_0^2\epsilon_l''\left |A_le^{ik_{zl}z'}+B_le^{-ik_{zl}z'}\right |^2= \nonumber \\
&\re \left[ k_{zl}(-A_l e^{ik_{zl}z}+B_l e^{-ik_{zl}z})(A_l e^{ik_{zl}z}+B_l e^{-ik_{zl}z})^*\right]\Big |_{z_{l}}^{z_{l-1}} \nonumber \\
&=Q_l(z_{l-1})-Q_l(z_{l}), 
\end{align}
which is valid for all layers. 
From the boundary condition, we have 
\begin{equation}
Q_l(z_{l})=Q_{l+1}(z_{l})
\end{equation}
Thus we find 
\begin{align}
&\sum \limits _{l=1}^{N+1} \int _{z_l}^{z_{l-1}} dz' k_0^2\epsilon_l''\left |A_le^{ik_{zl}z'}+B_le^{-ik_{zl}z'}\right |^2= \nonumber \\
&=Q_0(z_{0})-Q_{N+1}(z_{N+1}). 
\end{align}
For the last term, $z_{N+1}=-\infty$, so $Q_{N+1}(z_{N+1})=0$, and in our convention, $z_0=0$.  
The final result is 
\begin{align}
\re \left[ k_{z0}(1-r^s)(1+r^s)^*\right]= \nonumber \\
\left\{
\begin{gathered}
(1-|r^s|^2)|k_{z0}|,\qquad k_\rho<k_0 \\
2\im(r^s)|k_{z0}|,\qquad k_\rho>k_0 
\end{gathered} \right. . 
\end{align}

This is the contribution from s-polarized waves. For p-polarized waves, the corresponding identity is 
\begin{align}
\int _{z_l}^{z_{l-1}} dz' k_0^2\epsilon_l''\left (\left | \frac{k_{zl}(C_le^{ik_{zl}z'}-D_le^{-ik_{zl}z'})}{k_0\sqrt{\epsilon_l}}\right |^2 \right. \nonumber\\
\left.+\left |\frac{k_{\rho}(C_le^{ik_{zl}z'}+D_le^{-ik_{zl}z'})}{k_0\sqrt{\epsilon_l}}\right |^2 \right) \nonumber\\
=\re \left[ \frac{k_{zl}}{\sqrt{\epsilon_l}}(C_l e^{ik_{zl}z}-D_l e^{-ik_{zl}z}) \right. \nonumber\\
\left.(\sqrt{\epsilon_l}(C_l e^{ik_{zl}z}+D_l e^{-ik_{zl}z}))^*\right]\Big |_{z_{l}}^{z_{l-1}}
\end{align}

Then the contribution from p-polarized waves can be evaluated in the similar way. 
The final expression for thermal emission from a half space multilayered structure will be given by Eq.~\ref{general expression}. The reflection coefficients should be that of the whole structure. 

If we are interested in a slab inside vacuum rather than a half space structure, we can eliminate the contribution from the last layer vacuum part. To do so, in Eq.~(\ref{last layer}), for the last layer $A_{N+1}=0$ and $B_{N+1}=t^s$, the right hand side is therefore $\re (k_{z0}) |t^s|^2$, which vanishes for evanescent waves. Subtracting this term from Eq.~\ref{general expression} gives the thermal emission from a multilayered slab inside vacuum, 
\begin{align}
\label{thermal slab}
&u(\omega,z) = \frac{{U{}_{BB}(\omega ,T)}}{2} \nonumber \\
&\left\{ {\int_0^{{k_0}} {\frac{{{k_\rho }d{k_\rho }}}{{{k_0}\left| {{k_{z0}}} \right|}}} } \right.\frac{{(1 - |r^s|^2-|t^s|^2) +(1- |r^p|^2-|t^p|^2})}{2} \nonumber \\
& + \left. {\int_{{k_0}}^\infty  {\frac{{k_\rho ^3d{k_\rho }}}{{k_0^3\left| {{k_{z0}}} \right|}}{e^{ - 2\operatorname{Im} ({k_{z0}})z}}(\operatorname{Im} ({r^s}) + \operatorname{Im} ({r^p}))} } \right\}. 
\end{align} 
The above expression can be also obtained by replacing $ 1-|r|^2 $ in Eq.~\ref{general expression} with $ 1-|r|^2-|t|^2 $, which is consistent with Kirchoff's law. 

\section{Scattering matrix method and spatial coherence}
We now describe another approach to evaluating the near-field energy density near metamaterials using the scattering matrix approach. However, first we will discuss a few important points related to the concept of the thermal environment. We note that when the lower space is vacuum, the reflection coefficients are zero. As a result of Eq. (\ref{general expression}), the contribution from the evanescent waves part is zero while that from the propagating waves is nonzero. However, this is not very intuitive from FDT. The reason is that losses of vacuum i.e. $\epsilon''$ of vacuum is zero and from the second kind of FDT, the correlation function of random currents of vacuum should be zero, suggesting a zero field correlation. It turns out that for an unbounded vacuum region, we should add an infinitesimal imaginary part to $\epsilon_0$, integrate over the region and then take the limit of the imaginary part to be zero in the final expression\cite{landau_statistical_1980-1,kruger_trace_2012}. This is needed to preserve causality requirements. In the derivation of Eq.~(\ref{general expression}), we have integrated the source region $z'$ from $-\infty$ to $0$. However, for a vacuum gap with any finite width, the final fields correlation originating from the gap can be shown to be zero\cite{eckhardt_macroscopic_1984}. For this reason, fluctuations in vacuum can be interpreted to come from infinity. 

It is then natural to think about the thermal emission from the upper space vacuum region as well. If the vacuum region is also at temperature T, the system is at global thermal equilibrium. Therefore we can employ the first kind of FDT to calculate the thermal energy density. This approach is used in Ref.~\onlinecite{joulain_definition_2003} to define the local density of states. Here we directly cite the final result, 

\begin{align}
\label{total emission}
&u_{eq}(z,\omega ,T) = \frac{{U{}_{BB}(\omega ,T)}}{2} \nonumber \\
&\left\{ {\int_0^{{k_0}} {\frac{{{k_\rho }d{k_\rho }}}{{{k_0}\left| {{k_{z0}}} \right|}}} } \right.(2+\frac{k_\rho^2}{k_0^2}[\re(r^s e^{2ik_{z0}z})+\re(r^p e^{2ik_{z0}z})]) \nonumber \\
& + \left. {\int_{{k_0}}^\infty  {\frac{{k_\rho ^3d{k_\rho }}}{{k_0^3\left| {{k_{z0}}} \right|}}{e^{ - 2\operatorname{Im} ({k_{z0}})z}}(\operatorname{Im} ({r^s}) + \operatorname{Im} ({r^p}))} } \right\}
\end{align}

Note again that the contribution from evanescent waves equals that of Eq.~(\ref{general expression}), implying no evanescent waves contribution from the upper space vacuum region. 
However, in non-equilibrium, to determine electromagnetic fields induced by every random current inside the medium using second kind of FDT is quite laborious. We note from the second kind of FDT that the currents are not spatially correlated, which suggests that the thermal emission from different spatial regions can be calculated separately. In thermal equilibrium, we can calculate the thermal energy density by the first kind of FDT. Thus if we can calculate the thermal emission from the upper space vacuum part at temperature T, thermal emission only from the lower space can be achieved by excluding the vacuum part from the total thermal energy density. 

The electric field generated by the upper half vacuum space can be written as\cite{bimonte_general_2007} 
\begin{equation}
\vec E_f(\omega,\bs r)=\int \frac{d^2 \bs k_\perp}{4\pi^2} \vec E_f(\omega,\bs k_\perp, z) e^{i\bs k_\perp \cdot \bs r_\perp}
\end{equation}
where
\begin{equation}
\vec E_f(\omega,\bs k_\perp, z)=(a_s(\omega,\bs k_\perp)\hat s_-^0 + a_p(\omega,\bs k_\perp)\hat p_-^0 )e^{-ik_{z0}z}.
\end{equation}
$a_s$ and $a_p$ are the field amplitude for s and p-polarized waves, respectively. The operator $a=(a_s,a_p)^T$ satisfies the correlation function\cite{bimonte_general_2007}, 
\begin{align}
\left\langle {a(\omega,\bs k_\perp) \otimes a^\dagger (\omega',\bs k_\perp^\prime)} \right\rangle &=\nonumber\\
4\pi^2 C(\omega,\bs  k_\perp) &\delta(\omega-\omega')\delta^2(\bs k_\perp-\bs k_\perp^\prime). 
\end{align}

The coefficient $C$ can be read directly from FDT and the free space DGF, 
\begin{equation}
C(\omega,k_\perp)=\frac{\mu_0\omega}{4\pi }\Theta(\omega,T)\re{(\frac{1}{k_{z0}})}, 
\end{equation}
which vanishes for evanescent waves. 
These fluctuations from the upper vacuum region shines on the interface and get reflected. The total fields due to fluctuations in the vacuum part are
\begin{align}
\label{eq:vacuum fluctuation}
\vec E_0(z, & \omega,\bs k_\perp)=(a_s(\omega,\bs k_\perp)s_-^0 + a_p(\omega,\bs k_\perp)p_-^0 )e^{-ik_{z0}z} \nonumber \\
&+(r^s a_s(\omega,\bs k_\perp)s_+^0 + r^p a_p(\omega,\bs k_\perp)p_+^0 )e^{ik_{z0}z}.
\end{align}
The magnetic fields can be calculated using Eq.~(\ref{eq:vacuum fluctuation}) and Maxwell equations.  
Then one can find the energy density due to the fluctuations in the upper space vacuum, 
\begin{align}
\label{thermal vacuum}
& u_0(z,\omega,T)=\frac{{U{}_{BB}(\omega ,T)}}{2} \int_0^{k_0} \frac{{{k_\rho }d{k_\rho }}}{k_0|k_{z0}|} \Big \{1+  \nonumber \\
&  \frac{|r^s|^2+|r^p|^2}{2}+ \frac{k_\rho^2}{k_0^2}[\re(r^s e^{2ik_{z0}d})+\re(r^p e^{2ik_{z0}d})]\Big \}
\end{align}
Subtracting Eq.~(\ref{thermal vacuum}) from Eq.~(\ref{total emission}), we recover the expression by the second kind of FDT. 

From the definition of the cross-spectral density tensor
\begin{equation}
W(\bs r_1,\bs r_2, \omega)\delta(\omega-\omega')=\left\langle{\vec E(\bs r_1,\omega)\otimes \vec E^*(\bs r_2,\omega')}\right\rangle, 
\end{equation}
one can find the spatial coherence due to fluctuations in the upper space vacuum, 
\begin{align}
W_{zz}^0(\bs r_1,\bs r_2,\omega)=&\frac{{U{}_{BB}(\omega ,T)}}{4\epsilon_0} \int_0^{k_0} \frac{{k_\rho^3 d{k_\rho }}}{k_0^3|k_{z0}|} J_0(k_\rho d)\nonumber \\
&\big[\frac{1+|r^p|^2}{2}+\re(r^p e^{2ik_{z0}d})\big]
\end{align}
where $ \bs r_1=(0,0,z) $, $ \bs r_2=(d,0,z) $ and $ \int_{0}^{2\pi}d\theta e^{ik_\rho d \cos \theta}=2\pi J_0(k_\rho d) $ is used; $ J_0(k_\rho d) $ is the zeroth order Bessel function of the first kind.
Further, from Eq.~(\ref{first kind}), the first kind of FDT, we have 
\begin{align}
W_{zz}^{eq}(\bs r_1,\bs r_2,\omega)=& \frac{{U{}_{BB}(\omega ,T)}}{4\epsilon_0} \Big \{\int_0^{k_0} \frac{{k_\rho^3 d{k_\rho }}}{k_0^3 |k_{z0}|} J_0(k_\rho d) \nonumber \\
&\big[1+\re(r^p e^{2ik_{z0}d})\big]+ \int_{k_0}^{\infty} \frac{{k_\rho^3 d{k_\rho }}}{k_0^3 |k_{z0}|} \nonumber \\
& J_0(k_\rho d) \im(r^p)e^{-2\im(k_{z0})z} \Big \}
\end{align}
Then the contribution from the lower space structure is 
\begin{align}
W_{zz}(\bs r_1,\bs r_2,\omega)=& \frac{{U{}_{BB}(\omega ,T)}}{4\epsilon_0} \Big \{\int_0^{k_0} \frac{{k_\rho^3 d{k_\rho }}}{k_0^3 |k_{z0}|} J_0(k_\rho d) \nonumber \\
&\frac{1-|r^p|^2}{2}+ \int_{k_0}^{\infty} \frac{{k_\rho^3 d{k_\rho }}}{k_0^3 |k_{z0}|} \nonumber \\
& J_0(k_\rho d) \im(r^p)e^{-2\im(k_{z0})z} \Big \}
\end{align}
Only p-polarized waves contributes to $W_{zz}$ since s-polarized waves do not have $E_z$ components. 

Once again, if the structure is a multilayered slab in vacuum, the contribution from the lower vacuum space can be evaluated using the scattering matrix method in a similar way to the upper vacuum space. The fields due to the vacuum fluctuations in the lower space transmit through the planar structure,
\begin{equation}
\vec E_t(\omega,\bs k_\perp, z)=(t^s a_s(\omega,\bs k_\perp)\hat s_-^0 + t^p a_p(\omega,\bs k_\perp)\hat p_-^0 )e^{ik_{z0}z}.
\end{equation}
It is clear that the contributing energy density will be proportional to the $|t|^2$, so that we recover the result of Eq.~(\ref{thermal slab}). Note that due to the reciprocal property, the transmission coefficients from two sides of the structure should be identical. 

Generally speaking, considering a single object in thermal equilibrium, the energy density can be determined by the first kind of FDT, which is simply a single scattering event. To find the contribution from the object only, we can exclude the contribution from the environment, which can be also expressed by the scattering matrix of the object. If there are several objects at different temperatures, we can first decide the thermal emission from one specific object in the absence of other objects and then build the scattering part from other objects, in which procedure the temperatures of the other objects and the environment are assumed to be zero. Note this is the basic idea of M.Kardar and co-authors in sequent works\cite{kruger_nonequilibrium_2011,golyk_heat_2012,kruger_trace_2012}. Beyond the multilayered structures considered here, the authors also give the scattering matrix of various geometries including sphere and cylinder. For more complicated objects, numerical methods are also well developed.\cite{mccauley_modeling_2012,rodriguez_fluctuating-surface-current_2012,rodriguez_fluctuating-surface-current_2013,otey_fluctuational_2014}
 
\section{Results and discussions}
\begin{figure}
\includegraphics{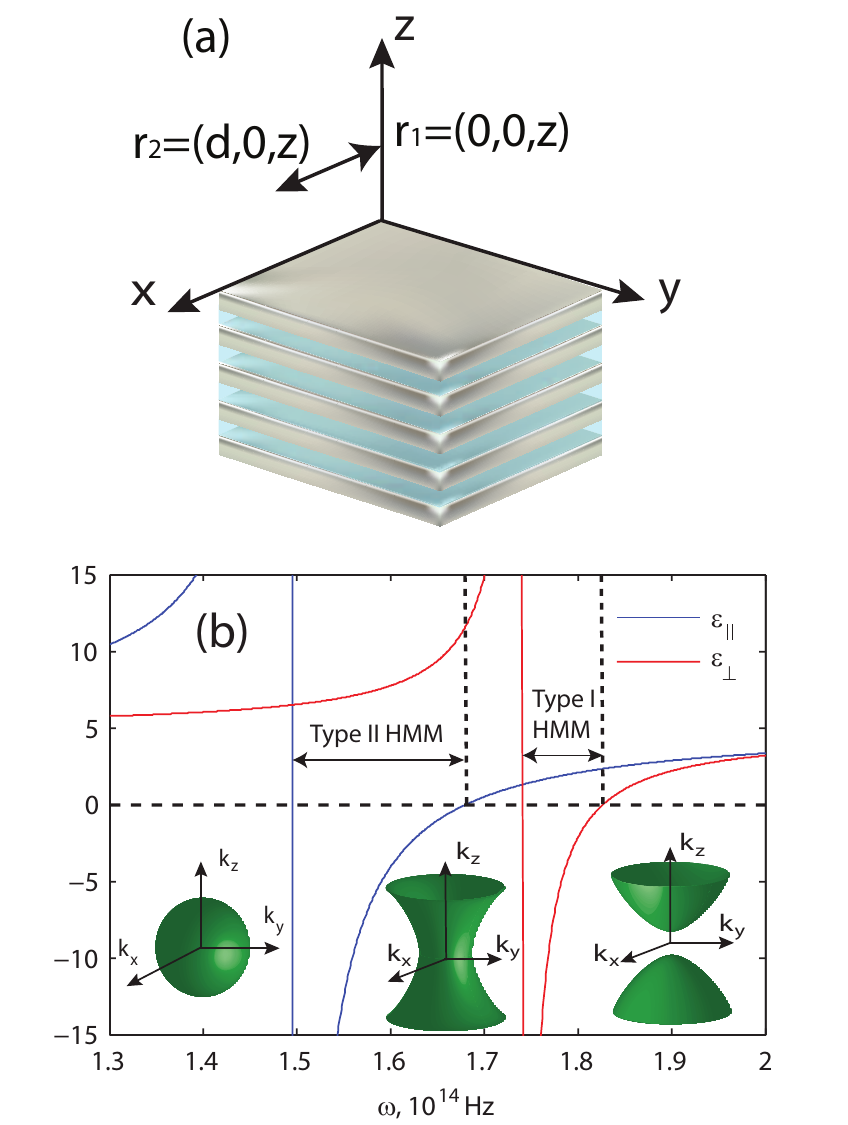}
\caption{\label{fig:emt_para} (a) Schematic of the multilayered structure and the coordinates. The spatial coherence are calculated between $\bs r_1=(0,0,z)$ and $\bs r_2=(d,0,z)$. (b) Effective permittivities of a SiO$_2$-SiC multilayered structure, where the fill fraction of SiC is 0.4. Only real part of the permittivity is plotted. The insets from left to right, denote the iso-frequency dispersion of dielectric, type II HMM and type I HMM.}
\end{figure}
\begin{figure}
\includegraphics{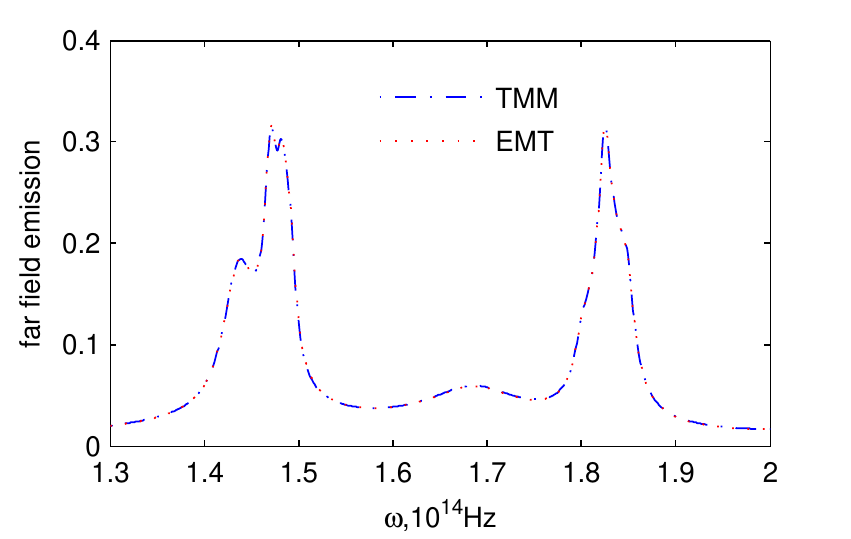}
\caption{\label{fig:far field} Normalized far field thermal emission of a 3$\mu$m SiO$_2$-SiC multilayered structure, with fill fraction of SiC is 0.4.}
\end{figure}
There are multiple approaches to achieving hyperbolic dispersion\cite{cortes_quantum_2012,guo_applications_2012}. Two of the prominent geometries consists of 1D or 2D periodic metal-dielectric structures. We consider here a multilayer combination of silicon dioxide (SiO$_2$) and silicon carbide (SiC) which has a metallic response in the Reststrahlen band due to phonon polaritons ($\re(\epsilon)<0$ between  $\omega_{TO}=149.5 \times 10^{12} $ Hz and $\omega_{LO}=182.7\times 10^{12}$ Hz, the transverse and longitudinal optical phonon resonance frequencies). The permittivity of SiC is given by $\epsilon_m=\epsilon_\infty(\omega_{LO}^2-\omega^2-i\gamma\omega)/(\omega_{TO}^2-\omega^2-i\gamma\omega)$, where $\omega$ is the frequency of operation, $\omega_\infty=6.7$ and $\gamma=0.9\times10^{12}$ Hz. We note that this realization formed the testbed for the first complete characterization of the modes of hyperbolic media due to their low loss as compared to plasmonic media\cite{korobkin_measurements_2010}. The modes of this HMM can be excited at relatively lower temperatures (400-500K) when the peak of black body emission lies within the Reststrahlen band of SiC.
\begin{figure}
\includegraphics{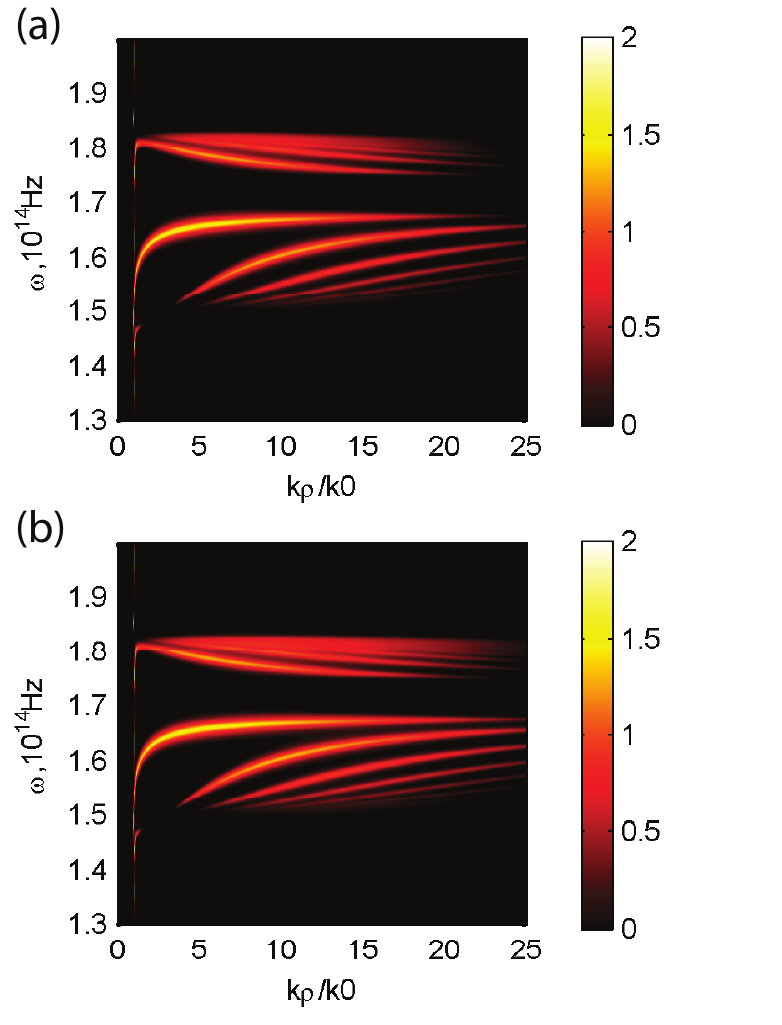}
\caption{\label{fig:1_EMT_dm} Wavevector resolved thermal emission (normalized to blackbody emission into the upper space) from a SiO$_2$-SiC multilayered structure calculated by (a) transfer matrix method and (b) EMT at z=200nm. The thermal emission is normalized to the black body emission to the upper half-space and in log scale. The structure consists of 40 layers of SiO$_2$/SiC , 30nm/20nm achieving a net thickness of 1$\mu$m. The presence of high-k modes are clearly evident in both the EMT calculation and the multilayer practical realization which takes into account all non-idealities due to dispersion, losses, finite unit cell size and finite sample size. The bright curves denote the enhanced thermal emission due to high-k modes in the HMM. In the practical multilayered structure, the high-k modes come from the coupled short range surface phonon polaritons at the silicon carbide and silicon dioxide interfaces.}
\end{figure}
To understand the thermal properties of phonon-polaritonic hyperbolic metamaterials we need to focus only on the Reststrahlen band of SiC where it is metallic. The multilayer structure (see schematic in Fig.~\ref{fig:emt_para}(a)) shows a host of different electromagnetic responses as predicted by effective medium theory $\epsilon_\parallel=\epsilon_mf+\epsilon_d(1-f)$ and $\epsilon_\perp=\epsilon_m\epsilon_d/(\epsilon_df+\epsilon_m(1-f))$, here $f$ is the fill fraction of the metallic medium\cite{cortes_quantum_2012}. 

We classify the effective uniaxial medium\cite{cortes_quantum_2012,guo_applications_2012} using the isofrequency surface of extraordinary waves which follow $k_z^2/{\epsilon_\parallel}+(k_x^2+k_y^2)/{\epsilon_\perp}=\omega ^2/c^2$  and the media are hyperboloidal only when $\epsilon_\parallel\epsilon_\perp<0$ . We can effectively achieve a type I hyperbolic metamaterial with only one negative component in the dielectric tensor ($\epsilon_\parallel>0$, $\epsilon_\perp<0$), type II hyperbolic metamaterial with two negative components ($\epsilon_\parallel<0$, $\epsilon_\perp>0$), effective anisotropic dielectric ($\epsilon_\parallel>0$, $\epsilon_\perp>0$) or effective anisotropic metal ($\epsilon_\parallel<0$, $\epsilon_\perp<0$). In Fig.~\ref{fig:emt_para}(b), we plot the effective permittivities of a SiO$_2$-SiC multilayered structure with the fill fraction 0.4 and label the two hyperbolic regions. As the purpose of this work is to examine how extraordinary waves in HMMs impact thermal emission properties, we only consider p-polarized waves in our numerical simulations. 

\subsection{Far field thermal emission}
We first characterize the thermal emission of a HMM slab in the far field. This is extremely important for experiments currently being pursued  in multiple groups.  We clearly observe two peaks in Fig.~\ref{fig:far field} in agreement with the previous work on epsilon-near-zero and epsilon-near-pole resonances for thermal emission\cite{molesky_high_2013}. The right one occurs when $\epsilon_\perp$ is close to zero. From the displacement field boundary condition, $\epsilon_{0}E_{0\perp}=\epsilon_{\perp}E_{1\perp}$, when $\epsilon_{\perp}\rightarrow 0$, the fields inside HMM $E_{1\perp}$ should be very large. Thus large absorption is expected at this epsilon near zero region.   The epsilon-near-pole resonance results in narrowband thermal emission due to the increase in the imaginary part of the dielectric constant in this ENP spectral  region. The most critical aspect is the direction of the dielectric tensor components which show ENZ or ENP\cite{molesky_high_2013}. An ENZ in the component parallel to the interface or an ENP perpendicular to the interface does not show such effects. 
 
\subsection{Near field thermal emission}
\begin{figure}
\includegraphics{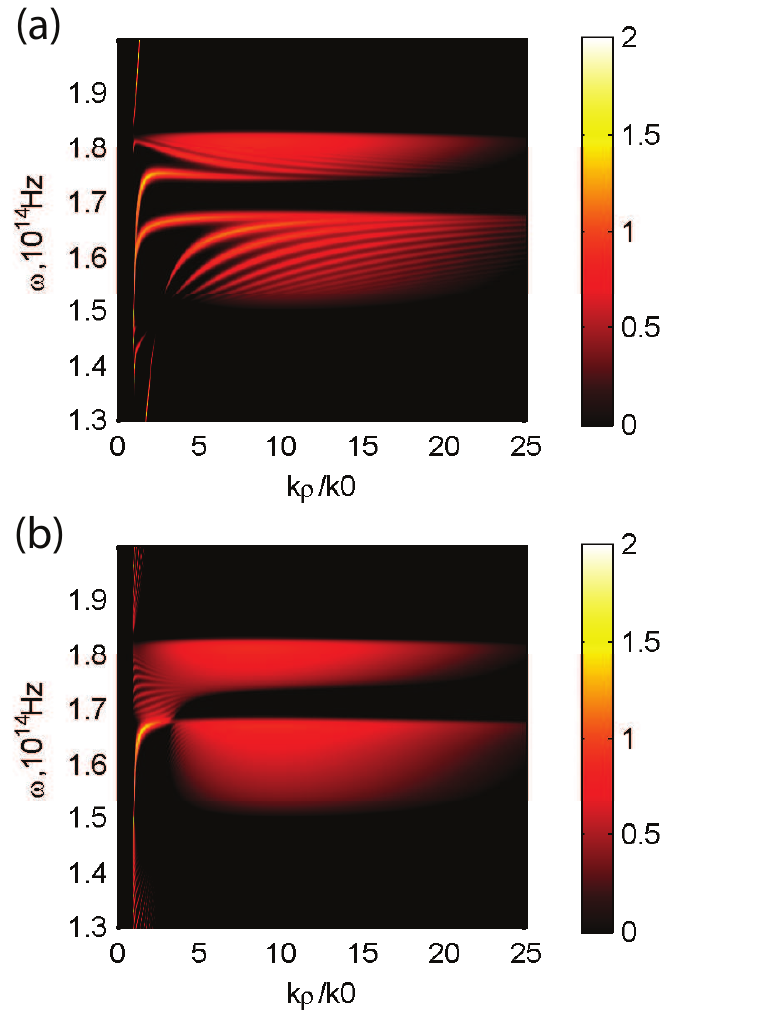}
\caption{\label{fig:3_30_EMT} Wavevector resolved thermal emission (normalized to blackbody emission into the upper space) from (a) a 3$\mu$m thickness HMM slab and (b) a 30$\mu$m thickness HMM slab. The fill fraction of SiC is 0.4, same as the 1$\mu$m HMM slab. The two hyperbolic regions where the thermal emission is enhanced are evident. The modes supported by 3$\mu$m thickness slab are denser than that of 1$\mu$m slab and the modes supported by the 30$\mu$m slab are almost continuous.}
\end{figure}

\begin{figure}
\includegraphics{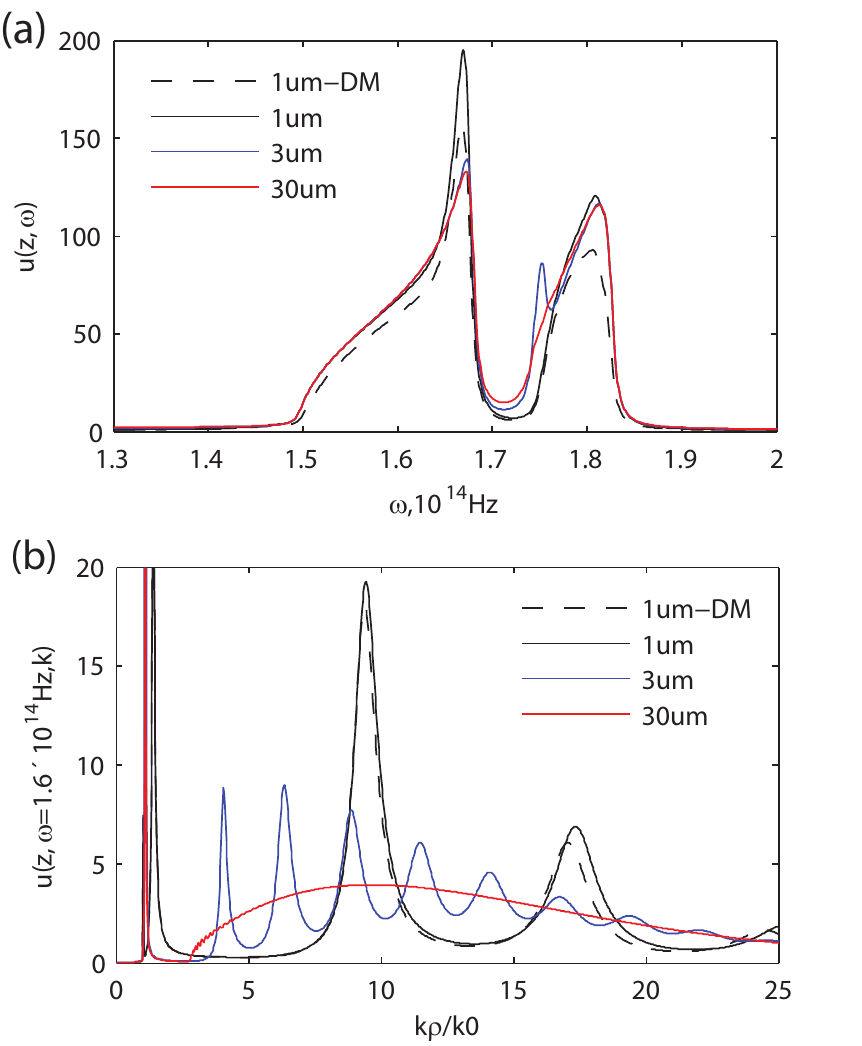}
\caption{\label{fig:emission} (a) Normalized thermal emission from slabs with various thicknesses. The dashed black line is calculated using transfer matrix method while the solid lines are calculated using EMT parameters, where 'DM' in the legend means the top layer of SiO$_2$(Dielectric)-SiC(Metal) multilayers is SiO$_2$. Despite the clear difference of the density of modes supported by the slabs shown in Fig.~\ref{fig:1_EMT_dm} and \ref{fig:3_30_EMT}, the thermal emission spectrum are interestingly in good agreement. The two main peaks where the thermal emission are largely enhanced are due to the high-k states in the two hyperbolic regions. (b) Wavevector resolved thermal emission at $\omega=1.6\times10^{14}$Hz. The sharp peaks on the left ($k_\rho/k_0<2 $) are the surface modes. When $k_\rho/k_0>3$, the curve for 30$\mu$m slab is almost flat with no oscillations, while that of 1$\mu$m and 3$\mu$m slabs show the discrete modes denoted by crests and troughs.}
\end{figure}

\begin{figure}
\includegraphics{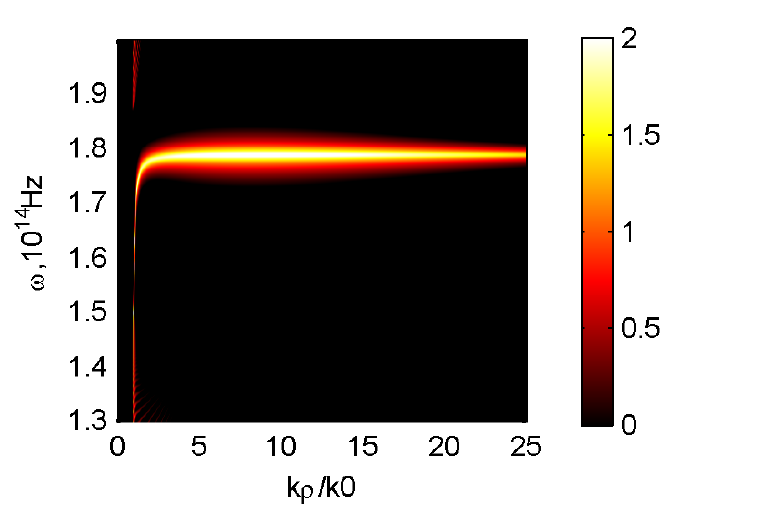}
\caption{\label{fig:SiC_phase} Thermal emission by a 30$\mu$m SiC slab. The red bright curve represents the dispersion of the SPhP mode between the vacuum and SiC interface since the slab is very thick.}
\end{figure}
\begin{figure}
\includegraphics{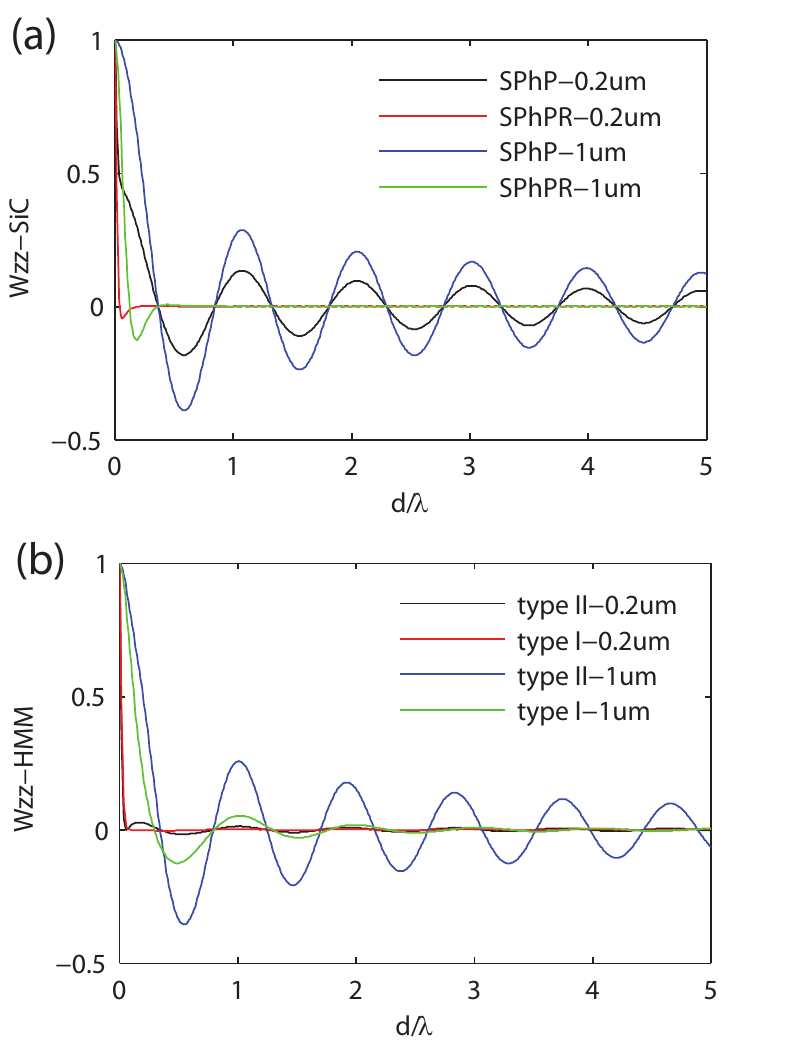}
\caption{\label{fig:spatial} spatial coherence of (a) a 30$\mu$m SiC slab and (b) a 30$\mu$m HMM slab at 0.2$\mu$m and 1$\mu$m from the surface with $\omega=1.6\times10^{14}$Hz and $\omega=1.79\times10^{14}$Hz. (a) At $\omega=1.6\times10^{14}$Hz, the SiC slab supports a single degenerate SPhP mode. As a result, SiC slab has large spatial coherence at both 0.2$\mu$m and 1$\mu$m. At $\omega=1.79\times10^{14}$Hz, the SPhP resonance frequency where $\re\epsilon_{\text{SiC}}=-1$, this frequency corresponds to a bright horizontal line in the SPhP dispersion curve shown in Fig.~\ref{fig:SiC_phase}. This means at this frequency, multi-modes with different wavevectors can be thermally excited. Thus the spatial coherence is poor both at 0.2$\mu$m and 1$\mu$m. (b)At $\omega=1.6\times10^{14}$Hz, the HMM slab supports high-k states besides the SPhP mode. At 0.2$\mu$m, the high-k states contribute a lot to the fluctuating electric fields, and consequently the spatial coherence is poor. But when the distance becomes larger at 1$\mu$m, the high-k states will not reach that far because of their large wavevector $k_\rho$. Thus the electric fields will be dominated by the surface mode which has smaller $k_\rho$. The spatial coherence length is large due to this dominant surface mode. At $\omega=1.79\times10^{14}$Hz, the HMM slab can only supports multiple high-k states, and unlike the type II HMM region, there is no lower bound for the high-k wavevectors. Thus the spatial coherence is poor both at 0.2$\mu$m and 1$\mu$m.}
\end{figure}

Here we analyze the near-field thermal emission from multilayer hyperbolic media\cite{guo_broadband_2012}. We first focus on how thermal emission will depend on the thickness of the slabs. In Fig.~\ref{fig:1_EMT_dm}, we plot the wavevector resolved thermal emission from a structure consists of 40 layers of SiO$_2$/SiC , 30nm/20nm achieving a net thickness of 1$\mu$m.  We clearly see multiple discrete high-k modes in both the type I and type II hyperbolic region. Note the thickness 1$\mu$m is about one tenth of the operating wavelength, so these high-k modes will not occur in conventional isotropic dielectrics. The excellent agreement between the EMT prediction and the practical multilayered structure is seen, which validates the use of EMT in our structure. Further, we increase the thickness of the slab to 3$\mu$m and 30$\mu$m while keeping the same unit cell. The waveguide modes will be denser as expected. At the thickness of 30$\mu$m, the high-k modes are almost continuous and result in two bright bands in Fig.~\ref{fig:3_30_EMT}(b). This is close to the bulk metamaterial limit.

We show the thermal emission spectrum in Fig.~\ref{fig:emission}(a) for various thicknesses of the metamaterial.   The two main peaks are due to the high-k modes in the hyperbolic region. In Fig.~\ref{fig:emission}(b), we plot  the wavevector resolved thermal emission at a specific frequency $\omega=1.6\times10^{14}$Hz within the type II hyperbolic region where the structure supports both surface mode and high-k modes. The sharp peaks at the left are due to the surface mode while the high-k modes emerge at larger $k_\rho$. In the high-k modes region, the curve for 30$\mu$m slab is almost flat indicative of a continuum of high-k modes. In contrast, the curves of 1$\mu$m and 3$\mu$m slabs clearly show the existence of discrete high-k waveguide modes featured by crests and troughs. 

\subsection{Spatial coherence of hyperbolic metamaterial slab}
Surface waves can lead to large spatial coherence length in the near field\cite{carminati_near-field_1999}. To see this, we first show in Fig.~\ref{fig:SiC_phase} the wavevector resolved thermal emission from a 30$\mu$m thick SiC slab. The bright curve gives the dispersion of surface phonon polariton (SPhP) between the vacuum and SiC interface. Note we will not see the splitting of the vacuum-SiC interface SPhP mode into long range and short range modes since 30$\mu$m is in the order of several operating wavelengths. In the time domain, the temporal coherence is best for monochromatic waves. Thus for the spatial coherence, one can imagine it will be favorable if a single wavevector dominates the fields among all the wavevectors. This is indeed the case for surface waves. In Fig.~\ref{fig:spatial}(a), we plot the spatial coherence of the SiC slab at $\omega=1.6\times 10^{14}$Hz and $\omega=1.79\times 10^{14}$Hz. At the frequency $\omega=1.6\times 10^{14}$Hz, the SPhP mode wavevector $k_\rho$ is about $1.1k_0$. Large spatial coherence length is seen at both 0.2$\mu$m and 1$\mu$m from the interface. However, near the surface phonon polariton resonance (SPhPR) frequency $\omega=1.79\times 10^{14}$Hz where $\epsilon_\text{SiC}=-1$, the mode dispersion curve is almost a horizontal line, which means that multiple modes with different wavevectors can be thermally excited. Thus a poor spatial coherence is expected. In Fig.~\ref{fig:spatial}(a), the spatial coherence is poor at at both 0.2$\mu$m and 1$\mu$m from the interface. This feature could be used to determine the resonance frequency.

Hyperbolic metamaterials can support multiple high-k modes. Therefore the spatial coherence length should not be long in the hyperbolic region. This is true for type I HMM. In Fig.~\ref{fig:spatial}(b), we plot $W_{zz}$ at $\omega=1.79\times 10^{14}$Hz, where the multilayered structure effectively behaves in the type I hyperbolic region. The spatial coherence lengths are only a fraction of the operating wavelength at both 0.2$\mu$m and 1$\mu$m from the interface.

But the situation for type II hyperbolic region is interestingly different. For a HMM slab in the type II hyperbolic region ($\epsilon_\parallel<0$, $\epsilon_\perp>0$), the slab can support a surface wave mode as well as multiple high-k modes. Thus we have two sets of modes that can result in a unique interplay of spatial coherence effects.  Furthermore, these modes are separated in wavevector space because of the lower bound of the high-k states in type II hyperbolic region\cite{cortes_quantum_2012}. High-k modes are confined to the surface better than suface waves and these high-k waves will dominate at a shorter distance from the interface. We choose $\omega=1.6\times 10^{14}$Hz within the type II hyperbolic region to confirm this point. At distance 0.2$\mu$m, the spatial coherence is very poor. However, at a larger distance 1$\mu$m, the fluctuating fields have large spatial coherence length. This is because at this distance, the contribution from surface wave mode dominates the electric fields while the high-k states rarely contribute to the fields. This distance dependence behavior can have applications such as obtaining the modes distribution at a given frequency. 

\subsection{Thermal Topological Transitions}
\begin{figure*}
\includegraphics{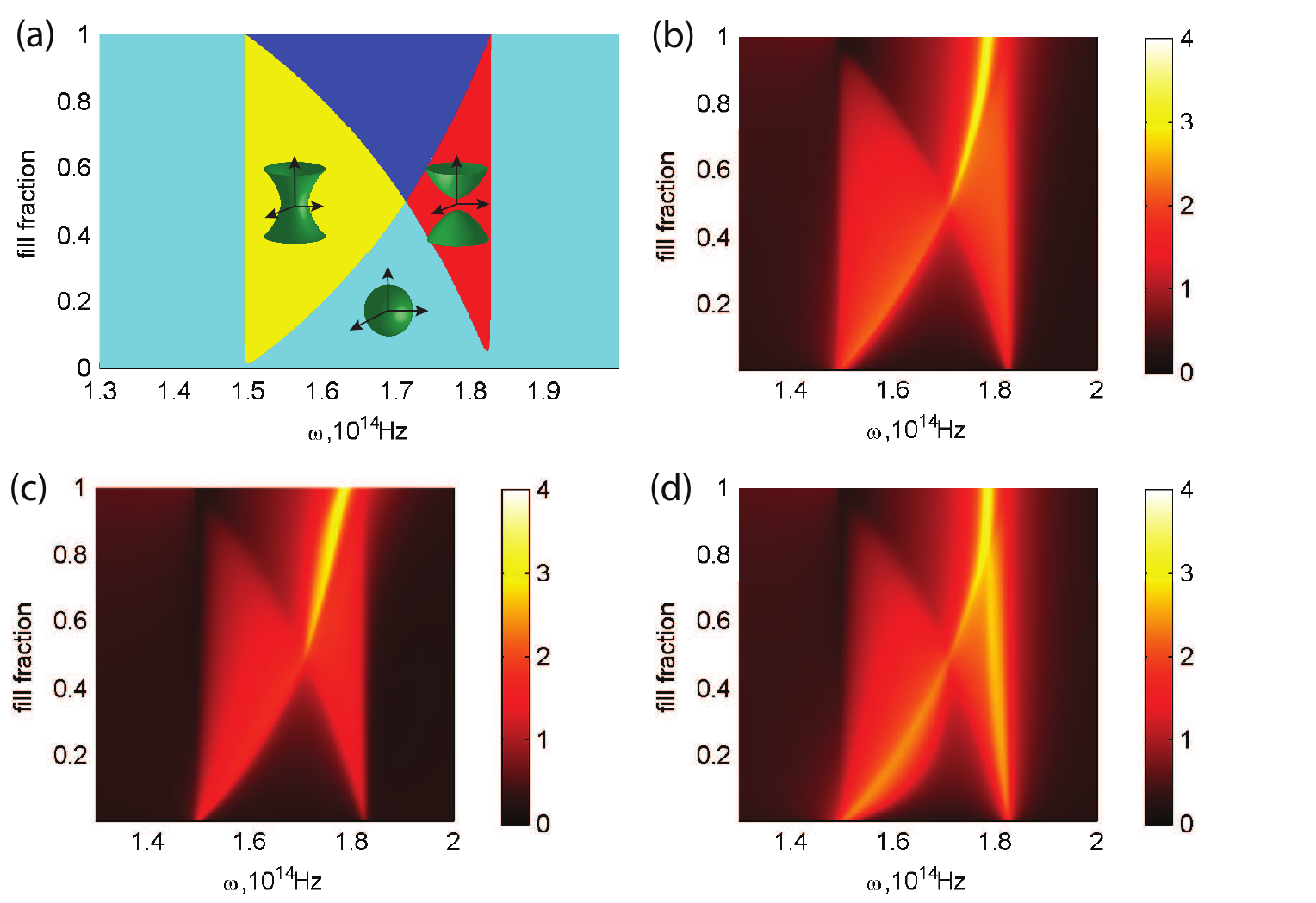}
\caption{\label{fig:TTT} (a) Optical phase diagram of SiC-SiO$_2$ multilayered structure predicted by EMT. Red region denotes effective dielectric, blue region means effective metal, yellow region stands for type I hyperbolic metamaterial, green region is type II hyperbolic metamaterial. Thermal emission at z=200nm (log scale plot normalized to the black body radiation into the upper half-space)  by the multilayered structure depending on the operating frequency and the fill fraction calculated by (b) EMT, (c) SiO$_2$-SiC multilayer (with first layer SiO$_2$), (d) SiC-SiO$_2$ multilayer (with first layer SiC). In the effective metal region, the dark red line is due to surface phonon polariton resonance. Both type I and type II region have a clear thermal emission enhancement due to bulk high-k modes  in agreement with the optical phase diagram.}
\end{figure*}
Until now, we have fixed the fill fraction to be 0.4. It is useful to examine the structure's behavior at various fill fractions. In Fig.~\ref{fig:TTT}(a), we plot the optical phase diagram\cite{cortes_quantum_2012,guo_broadband_2012} of this metamaterial which shows the  isofrequency surfaces achieved at different frequencies and fill fractions of SiC. The phase diagram is classified as effective dielectric, effective metal, type I and type II HMM as introduced before\cite{cortes_quantum_2012,guo_applications_2012}. 

Figure.~\ref{fig:TTT}(b) shows the thermal energy density (normalized to black body radiation into the upper half space) evaluated using Rytov's fluctuational electrodynamics for an effective medium slab at a distance of z=200nm from the metamaterial. It is seen that the regions of hyperbolic behavior exhibit super-Planckian thermal emission in agreement with our previous analytical approximation, but here we will go beyond effective medium theory and consider practical structures. The role of the surface waves is very important and can lead to significant deviations when the unit cell size is not significantly subwavelength\cite{kidwai_effective-medium_2012,tschikin_limits_2013,guo_thermal_2013}. 

The macroscopic homogenization utilized to define a bulk electromagnetic response is valid when the wavelength of operation exceeds the unit cell size ($\lambda\gg a$). However, even at such wavelengths if one considers incident evanescent waves on the metamaterial the unit cell microstructure causes significant deviations from EMT. This is an important issue to be considered for quantum and thermal applications where the near-field properties essentially arise from evanescent wave engineering (high-k modes)\cite{cortes_quantum_2012,guo_applications_2012}. For the multilayer HMM, at distances below the unit cell size, the thermal emission is dominated by evanescent waves with lateral wavevectors $k_\rho\gg1/a$. Since this is above the unit-cell cut off of the metamaterial, the high-k modes do not contribute to thermal emission at such distances. It is therefore necessary to consider thermal emission from a practical multi-layer structure taking into account the layer thicknesses. This is shown in Fig.~\ref{fig:TTT}(c) and Fig.~\ref{fig:TTT}(d). The unit cell size is 200nm, and we consider a semi-infinite multilayer medium using the formalism outlined in Ref.~\onlinecite{kidwai_effective-medium_2012}. An excellent agreement is seen of the optical phases of the multilayer structure with the EMT calculation.

\section{conclusion}
This work shows that extension of equilibrium and non-equilibrium fluctuational electrodynamics to the case of metamaterials can lead to novel phenomena and applications in thermal photonics. 
We presented a unified picture of far-field and near-field spectra for experimentalists and also introduced the near-field spatial coherence properties of hyperbolic media. We have analyzed in detail thermal topological transtions and super-Planckian thermal emission in practical phonon-polaritonic hyperbolic metamaterials. We paid particular attention not only to the effective medium approximation but discussed all non-idealities limiting the super-planckian thermal emission from HMMs.  We have provided practical designs to experimentally measure and isolate our predicted effect. Our work should lead to a class of thermal engineering applications of metamaterials. 

\section{acknowledgment}
Z. Jacob wishes to acknowledge discussions with E. E. Narimanov. This work was partially supported by funding from Helmholtz Alberta Initiative, Alberta Innovates Technology Futures and National Science and Engineering Research Council of Canada.


%

\end{document}